\documentclass[reprint,aps,pra,superscriptaddress]{revtex4-1}
\usepackage{natbib}
\usepackage{amsmath}
\usepackage{amssymb}
\usepackage{amsthm}
\usepackage{nicefrac}
\usepackage{graphicx}
\usepackage{bm}

\newcommand{\im}[1]{\operatorname{Im}\left[#1\right]}
\newcommand{\re}[1]{\operatorname{Re}\left[#1\right]}

\newcommand{\sym}[1]{\operatorname{Sym}\left[#1\right]}
\newcommand{\asym}[1]{\operatorname{Asym}\left[#1\right]}

\begin{document}
\title{Hierarchical Mean-Field $\mathbb{T}$ Operator 
Bounds on Electromagnetic Scattering: \\
Upper Bounds on Near-Field Radiative Purcell Enhancement}
\author{Sean Molesky}
\thanks{Equal contribution}
\author{Pengning Chao}
\thanks{Equal contribution}
\author{Alejandro W. Rodriguez}
\affiliation{Department of Electrical Engineering, Princeton 
University, Princeton, New Jersey 08544, USA}
\email{arod@princeton.edu}
\begin{abstract}
  We show how the central equality of scattering theory---the
  definition of the $\mathbb{T}$ operator---can be used to generate
  hierarchies of mean-field constraints that act as natural
  complements to the standard electromagnetic design problem of
  optimizing some objective with respect to structural degrees of
  freedom. Proof-of-concept application to the problem of maximizing
  radiative Purcell enhancement for a dipolar current source in the 
  vicinity of a structured medium, an effect central to many sensing 
  and quantum technologies, yields performance bounds that are 
  frequently more than an order of magnitude tighter than all 
  current frameworks, highlighting the irreality of these models in 
  the presence of differing domain and field-localization length 
  scales.
  Closely related to domain decomposition and
  multi-grid methods, similar constructions are possible in any 
  branch of wave physics, paving the way for systematic evaluations 
  of fundamental limits beyond electromagnetism.
\end{abstract}
\maketitle

Accelerating over the last decade, the adoption of inverse design
techniques like ``density'' (``topology'') or level-set optimization
in photonics~\cite{jensen2011topology,molesky2018inverse}---ideally 
matching \emph{structural degrees of freedom} to the computational 
grid---has vastly simplified the challenge of 
discovering geometries with remarkable optical characteristics, 
leading to improved designs for wide band-gap photonic 
crystals~\cite{men2014robust,meng2018topology,chen2019inverse}, 
enhanced polarization control~\cite{okoro2017demonstration,
callewaert2018inverse}, ultra-thin optical elements~\cite{meem2020large,chung2020high,
lin2020end}, and topological materials~\cite{lin2016enhanced,
long2019inverse,christiansen2019topological}. 
However, because navigating the immense range of allowed structures 
in such formulations necessitates reliance on local information 
(approximations based on function evaluations, gradients, 
etc.~\cite{gill2019practical}) and the relation between fields and 
structural variations set by Maxwell's equations is 
non-convex~\cite{angeris2019computational}, it is rarely known how
close these solutions are to true (global) optima, or to what extent
they are determined by design choices (e.g. system length scales,
material susceptibility, and properties of the algorithm) as opposed
to physical principles.

Recently, a number of promising proposals for addressing these
knowledge gaps have been put forward by combining Lagrange duality
with physical consequences of scattering theory~\cite{
  molesky2020t,gustafsson2019upper,kuang2020maximal,
  trivedi2020fundamental}---relaxing the true \emph{local} constraints
of wave physics to \emph{global} conservation principles, Fig.~1.  And
in particular, exploiting an optical theorem requiring that real and
reactive power be conserved \emph{on
  average}~\cite{jackson1999classical, gustafsson2019upper}, has been
shown to produce performance limits for propagating waves that
accurately anticipate the results of density optimization (usually
within factors of unity) across a variety of
examples~\cite{molesky2020t}.  Yet, when these same techniques are
applied to situations where evanescent (near-field) wave effects
dominate overall behavior, Fig.~2, calculated bounds and objective
values obtained by density optimization often differ by several orders
of magnitude, and exhibit markedly different trends.

\begin{figure*}[t!]
\centering
\includegraphics[width=2.0\columnwidth]{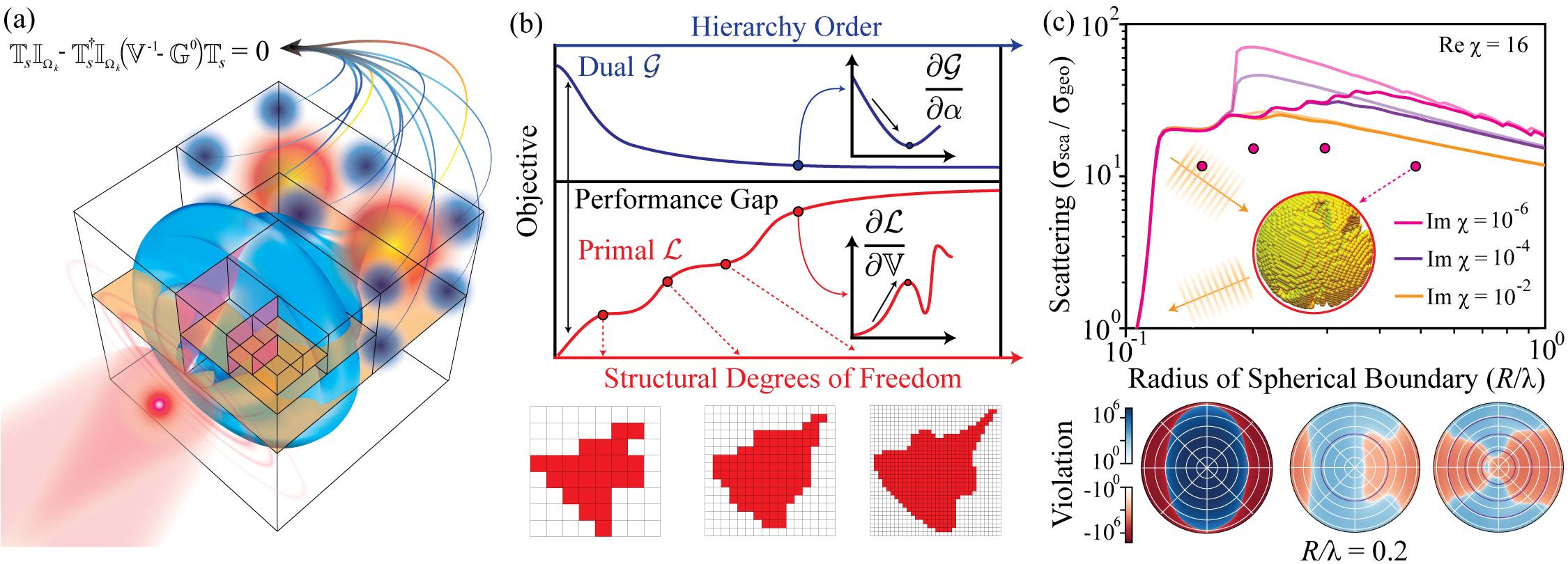}
\vspace{-10 pt}
\caption{\textbf{Schematic mean-field hierarchical bounds and
    application to scattering cross sections.} (a) Like Maxwell's
  equations, the definition of the $\mathbb{T}$ operator,
  \eqref{Tformal}, implicitly contains relations that must be obeyed
  at every spatial point. Present approaches to electromagnetic
  limits, however, only impose that this equality be satisfied on
  average, allowing bound solutions to exhibit highly unphysical local
  features. (b) By enforcing that the definition of the $\mathbb{T}$
  operator be respected on successively smaller subdomains or
  clusters, described by constraints $\alpha$, the dual bound
  objective $\mathcal{G}$ given in \eqref{gSol} acquires a
  hierarchical structure that mirrors the primal problem of optimizing
  some objective $\mathcal{L}$ in \eqref{Lgen}, in terms of an
  increasing number of structural degrees of freedom contained in the
  scattering potential $\mathbb{V}$.  In the limit of point clusters
  and complete structural freedom, the two problem statements are
  equivalent; see discussion in main text.  (c) Upper bounds on the
  plane-wave scattering cross-section $\sigma_\text{sca}$, relative to
  the geometric cross section of the domain $\sigma_\text{geo}$, for
  any structure bounded by a sphere of radius $R$. Lighter lines
  result by enforcing that power be conserved globally, as in
  Refs.~\cite{molesky2020t, gustafsson2019upper}. Similarly colored
  dark lines result when analogous equalities are asserted over eight,
  evenly spaced, radial subdomains. A profile of one of the
  density-optimized structures is shown as an inset. Shown below are
  logarithmic color maps of the corresponding violation in the real
  part of \eqref{regT} for one, two and four (evenly spaced) shell
  clusters, in the plane perpendicular to both the incoming wave
  vector and direction of polarization.}
\vspace{-10 pt}
\end{figure*}

In this Letter, we remedy this issue, and euclidate fundamental
connections between dual bounds and structural optimization, by
proposing the adoption of constraint hierarchies originating from the
central equality of scattering theory---the definition of the
$\mathbb{T}$ operator. 
The approach functions, in essence, as a collection of successively 
refined mean-field theories~\cite{opper2001advanced}. 
At the base of the hierarchy, only the global (spatially integrated) 
real and reactive power constraints studied in 
Refs.~\cite{molesky2020t,gustafsson2019upper} are imposed on the 
optimization objective, equating these prior results to a 
first-order mean-field approximation. 
In every subsequent refinement, the computational
domain is decomposed (partitioned) into increasingly small, nested,
subdomains, which, through projection, induces additional scattering
constraints and results in a higher-order approximation, Fig.~1.
Reminiscent of multi-grid and multi-scale 
methods~\cite{chen2012discontinuous,zhu2006multigrid}, the order of
the hierarchy thus acts as a resolution ``knob'' for systematically
controlling the extent to which Maxwell's equations are respected
\emph{locally}, allowing multiple length scales (beyond the size of
the domain) to be considered concurrently. 
Correspondingly, the method also presents a complementary top-down 
approach to inverse design: the solution of any optimization problem 
in the constraint hierarchy is always ``more optimal'' than what is 
conceptually possible if full wave physics are completely obeyed, 
whereas inverse design in a finite number of structural degrees of 
freedom is always suboptimal; in the limit of point subdomains 
(infinite mean-field order) and infinitesimal structural variations 
(vanishing ``voxels'') the two views agree, and, if strong duality
holds~\cite{boyd2004convex}, the bounds solution in fact determines a
globally optimal structure. 
More concretely, proof-of-concept application of the method to the 
problem of enhancing radiative emission from a dipolar current 
source in the near field of a structured medium, crucial to 
optical sensing and quantum information technologies~\cite{
raussendorf2001one,belushkin2018nanoparticle,
chakravarthi2020inverse}, yields bounds that come substantially closer to the values found by density optimization. 

\textit{Notation}---Throughout, $\mathbb{I}$ is used to denote the 
identity operator, and subscripts on operators (blackboard bold 
letters) are used as a booking device for the domains and codomains 
of definition. 
When only a single subscript is shown, the domain and codomain are 
identical. 
The subscripts $b$ and $s$ mark spatial locations as part of either 
the background ($b$) or scattering object ($s$) within in some 
predefined domain, $\Omega$. 
When an operator appears without subscripts, its domain and codomain 
are $\Omega$. 
$\mathbb{G}^{\text{0}}$ refers to the background Green's 
function~\cite{molesky2020t}, and hence depends on $\Omega$.
$\mathbb{V}$ is used to denote the scattering potential, i.e. any
polarizable medium not included in $\mathbb{G}^{\text{0}}$.

\textit{Constraints}---In scattering
theory~\cite{tsang2004scattering,molesky2020fundamental}, the role
commonly played by wave equations (e.g. Maxwell's equations) is
typified by the $\mathbb{T}$ operator, defined as
\begin{equation}
  \mathbb{I}_{s} = \mathbb{I}_{s}
  \left(\mathbb{V}^{-1} - \mathbb{G}^{\text{0}}\right)
  \mathbb{T}_{s};
  \label{Tformal}
\end{equation}
which, together with knowledge of $\mathbb{G}^{\text{0}}$ and
$\mathbb{V}$, abstractly determines all fields for a given source.
This operational picture lies at the heart of the hierarchical
construction. 
Any number of manifestly true relations can be generated by probing 
\eqref{Tformal} with linearly independent combinations of fields 
from the right and linear functionals from the left, and, by 
Lagrange duality, any set of constraints can be used to produce 
calculable bounds on any given optimization objective (so long as 
the dual remains soluble)~\cite{boyd2004convex}. 
Therefore, because every relation derived from \eqref{Tformal} is 
physical, each and every such collection of equalities generates 
some physical bound on any wave process.

As may be expected, certain choices are more naturally motivated than
others, and the difficulty of the associated convex optimization
problem that must be solved in each case depends closely on the
particular constraints chosen. 
If all identities contained in \eqref{Tformal} are imposed over some 
complete basis, then the associated primal problem is equivalent to 
completely free structural optimization, and in computing a solution 
to the dual (convex) system an actual $\mathbb{T}$ operator must be 
nearly constructed. 
(The collection of all input--output relationships determines any 
operator, and so, the only caveat that makes this statement inexact 
is that the solution of the dual problem may not satisfy every 
constraint.)
Hence, making full use of \eqref{Tformal} likely results in an
optimization problem comparable to bottom-up structural inverse
design~\cite{lazarov2016length,jin2020inverse}. 
On the other hand, if only a select subset of the information 
contained in \eqref{Tformal} is kept, the simplicity of determining 
bounds can be greatly reduced, at the cost of allowing the
discovered solution to inevitably violate some local scattering
relation(s). 
But, unlike standard calculations where the implications of some 
``partially valid'' model on actual solutions are seldom known a 
priori, solving a dual problem always determines a bound on
performance, and therefore always yields useful information.

To begin, it is simplest to work with \eqref{Tformal} using either 
the image or \emph{polarization field} resulting from the action of 
$\mathbb{T}_{s}$ on a single source field $\left|\textbf{S}\right>$,
$\mathbb{T}_{s} \left|\textbf{S}\right> \mapsto
\left|\textbf{T}\right>$~\cite{molesky2020t}, or, with this image and
the action of $\mathbb{I}_{s}$ on $\left|\textbf{S}\right>$,
$\left\{\mathbb{T}_{s}\left|\textbf{S}\right> \mapsto \left|
\textbf{T}\right>,~\mathbb{I}_{s}\left|\textbf{S}\right> \mapsto
\left|\textbf{R}\right>\right\}$~\footnote{Additional sources and
  fields, not necessarily related to the problem, may also be 
  included to more completely explore $\mathbb{T}$, and push the 
  resulting primal optimization problem towards structural inverse 
  design.}.
Letting $\mathcal{R} = \left\{\Omega_{k}\right\}_{_{K}}$ denote the
sets of chosen subdomains, the first choice leads to \emph{cluster}
constraints of the form
\begin{align}
  &\left(\forall \Omega_{k}\in\mathcal{R}\right)~
  \left<\textbf{S}\right|\mathbb{I}_{_{\Omega_{k}}}
  \left|\textbf{T}\right> = 
  \left<\textbf{T}\right|\mathbb{U}\mathbb{I}_{_{\Omega_{k}}}
  \left|\textbf{T}\right>,
  \label{regT}
\end{align}
where $\mathbb{U} =
\mathbb{V}^{-1\dagger}-\mathbb{G}^{\text{0}\dagger}$ so that
$\asym{\mathbb{U}}$ is positive definite, and $\langle \textbf{F} |
\textbf{G}\rangle = \int d^3\textbf{x}\, \textbf{F}(\textbf{x})^*
\cdot \textbf{G}(\textbf{x})$ is a complex-conjugate inner product
(spatial integration over the entire domain). 
The second choice allows for greater variety, and, $\left(\forall
\Omega_{k}\in\mathcal{R}\right)$, any combination of 
\begin{align}
  &\left<\textbf{S}\right|\mathbb{I}_{_{\Omega_{k}}}
  \left|\textbf{R}\right> = 
  \left<\textbf{R}\right|\mathbb{I}_{_{\Omega_{k}}}
  \left|\textbf{R}\right>,
  &\left<\textbf{S}\right|\mathbb{I}_{_{\Omega_{k}}}
  \left|\textbf{T}\right> &=
  \left<\textbf{R}\right|\mathbb{I}_{_{\Omega_{k}}}
  \left|\textbf{T}\right>,
  \nonumber \\
  &\left<\textbf{S}\right|\mathbb{I}_{_{\Omega_{k}}}
  \left|\textbf{R}\right> = 
  \left<\textbf{T}\right|\mathbb{U}\mathbb{I}_{_{\Omega_{k}}}
  \left|\textbf{R}\right>,
  \label{regTR}
\end{align}
appear to be, at least presently, sensible.

Both \eqref{regT} and \eqref{regTR} follow from \eqref{Tformal} 
based on the properties of $\mathbb{I}_{s}$, $\mathbb{T}_{s}$, and 
the commutativity of spatial projection. 
$\left(\forall~U,~V\subset\mathbb{R}^{n}\right)~U\cap V = V\cap U$, 
and so, as $\mathbb{I}_{s}$ and $\mathbb{I}_{_{\Omega_{j}}}$ both 
denote projections into spatial locations, 
$\mathbb{I}_{_{\Omega_{j}}}\mathbb{I}_{s} = 
\mathbb{I}_{s\cap_{\Omega_{j}}} = 
\mathbb{I}_{s}\mathbb{I}_{_{\Omega_{j}}}$, implying that (for any 
$\Omega_{j}\subset\Omega$) $\mathbb{I}_{s}
\mathbb{I}_{_{\Omega_{j}}} = \mathbb{I}_{s}^{2}
\mathbb{I}_{_{\Omega_{j}}} = \mathbb{I}_{s}
\mathbb{I}_{_{\Omega_{j}}}\mathbb{I}_{s}$, 
$\mathbb{I}_{_{\Omega_{j}}}\mathbb{T}_{s} = 
\mathbb{I}_{_{\Omega_{j}}}\mathbb{I}_{s}\mathbb{T}_{s} = 
\mathbb{I}_{s}\mathbb{I}_{_{\Omega_{j}}}\mathbb{T}_{s}$, and 
$\mathbb{I}_{_{\Omega_{j}}}\mathbb{I}_{s} = 
\mathbb{I}_{_{\Omega_{j}}}\mathbb{T}_{s}^{\dagger}\mathbb{U}
\mathbb{I}_{s} = \mathbb{T}_{s}^{\dagger}\mathbb{U}
\mathbb{I}_{_{\Omega_{j}}}\mathbb{I}_{s}$.
If only a single cluster corresponding to the entire domain $\Omega$ 
is used, then \eqref{regT} reduces to the constraints examined in 
Ref.~\cite{molesky2020t}, with the symmetric (Hermitian) and 
anti-symmetric (skew-Hermitian) parts of \eqref{regT}, 
$\im{\left<\textbf{S}|\textbf{T}\right>} = 
\left<\textbf{T}\right|\asym{\mathbb{U}}\left|\textbf{T}\right>$ and 
$\re{\left<\textbf{S}|\textbf{T}\right>} = \left<\textbf{T}\right|
\sym{\mathbb{U}}\left|\textbf{T}\right>,$ representing the global 
(averaged over the entire scatterer) conservation of real and 
reactive power, respectively.

Every equality of the form given by \eqref{regT} represents a similar
requirement on how power may be transferred between an exciting field
and the response (polarization) current it generates. 
As verified in Fig.~2, these additions are crucial for properly 
describing near-field interactions. 
When no constraints other than the global conservation of real and 
reactive power are imposed, the optimal $\left|\textbf{T}\right>$ 
discovered via the method presented in Refs.~\cite{molesky2020t,
gustafsson2019upper} conserves power only by canceling equally large 
positive and negative violations over $\Omega$, Fig.~1: locally, the 
power drawn from the incident field is either far greater, or far 
smaller, than what can be accounted for by absorption and 
scattering. 
By successively subdividing the domain, the scale over which such 
cancellations of unphysical behavior can occur is continually 
reduced, and because there is always an implicit interaction length 
scale in $\mathbb{U}$ set by material loss and the background 
Green's function, these tighter requirements ultimately lead to 
increasingly physical fields. 
Intuitively, beyond a certain critical size, the effect produced by 
rapid spatial variations of violation is no different than an 
averaged ``mean'' field which respects the constraints at all 
spatial points.

Though notionally similar, the role of \eqref{Tformal} in 
\eqref{regTR} (third line) requires a distinct interpretation. 
Rather than describing how power may be transferred between the 
source and the polarization field it generates within subregions of 
the domain, these relations state that within the scatterer 
$\mathbb{U}^{\dagger}$ must effectively ``invert'' 
$\left|\textbf{T} \right>$, reproducing the projection of the source 
into the scattering object, $\left|\textbf{R}\right>$. 
If \eqref{regTR} were true pointwise, instead of on average over 
some finite collection of subsets, then both $\left|\textbf{R}
\right>$ and $\left|\textbf{T}\right>$ would be produced by a 
geometry with scattering material at all locations
$\textbf{x}$ where $\textbf{R}(\textbf{x}) = \textbf{S}(\textbf{x})$,
equating this limit with structural inverse design.

\begin{figure*}[t!]
\centering
\includegraphics[width=2.0\columnwidth]{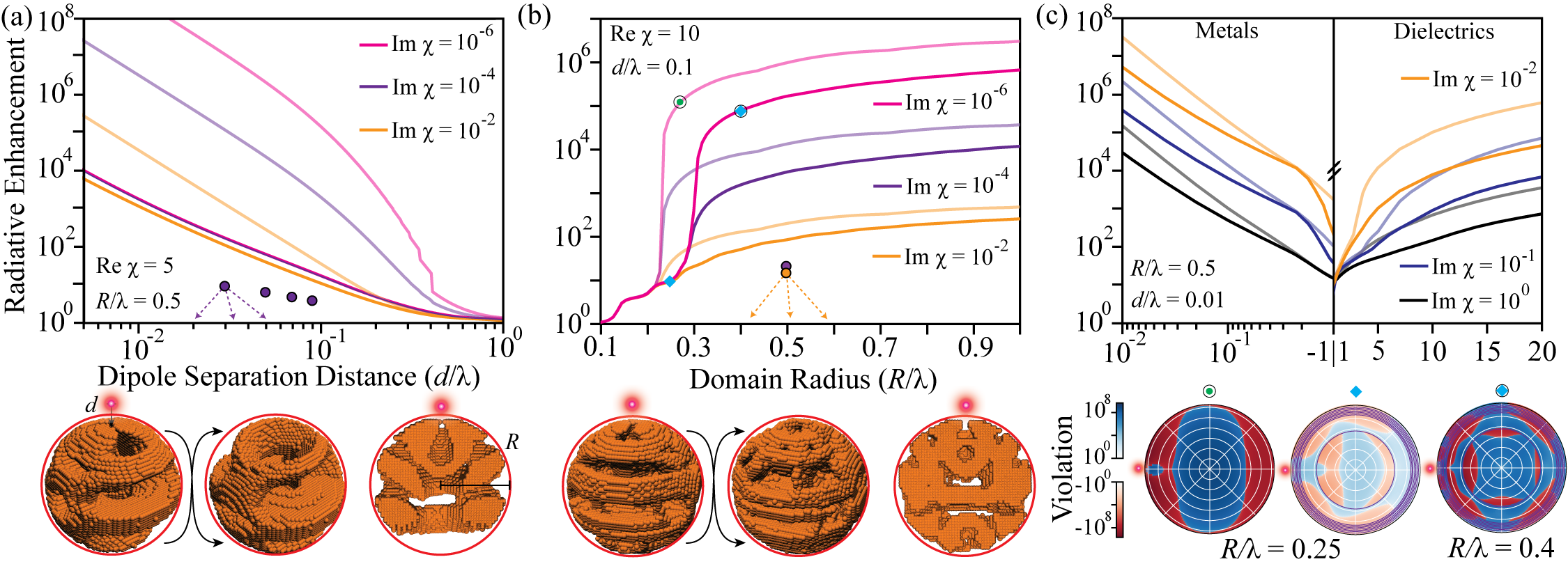}
\vspace{-10 pt}
\caption{\textbf{Application to radiative Purcell enhancement.}
  Following the same conventions as Fig.~1(c), for a dipolar
  source, proof-of-concept radial shell cluster constraint 
  hierarchies are found to substantially alter both the conditions 
  under which resonant (loss dependent) radiative emission 
  enhancement can possibly occur, panels (a) and (b), and generally 
  reduce the dependence of calculated limits on material properties, 
  all panels. 
  Throughout, the solid lines are calculated with 
  $r_{\text{shell}}/R =
  \left\{0.97,0.95,0.93,0.91,0.89,0.87,0.85,0.6\right\}$. 
  Key features are described in the main text. 
  Profiles and cross cuts of two representative density optimized 
  structures are displayed below (a) and (b). 
  The color maps below (c) show spatial violations in the real part 
  of \eqref{regT} for the parameter values marked in (b). 
  No notable Purcell enhancement is observed for $-1 < \text{Re}
  \left[\chi\right]< 1$.}
\vspace{-10 pt}
\end{figure*}

\textit{Hierarchy}---Through \eqref{regT} and \eqref{regTR}, any set
of regions in $\Omega$ defines a collection of constraints on any
observable property of the associated scattering theory; and in turn,
these constraints define an optimization problem that bounds the
observable, along with a dual solution $\left|\textbf{T}_{d}\right>$
or $\left\{\left|\textbf{T}_{d}\right>, ~
\left|\textbf{R}_{d}\right>\right\}$. 
Consider the collection $\mathcal{C}$ of all such sets of regions, 
$\mathcal{R} = \left\{\Omega_{k}\right\}_{K}$, with the properties 
that $\bigcup_{K}\Omega_{k} = \Omega$ and $\left(\forall k\neq
j\right)~\Omega_{k}\cap\Omega_{j} = \emptyset$. 
Call $\mathcal{R}'$ a refinement of $\mathcal{R}$, $\mathcal{R}'\geq 
\mathcal{R}$, if $\left(\forall \Omega_{k}\in\mathcal{R}\right)~
\exists\left\{\Omega_{j}\right\}_{J}\subset\mathcal{R}'\ni \Omega_{k}
= \bigcup_{j\in J}\Omega_{j}$, giving $\mathcal{C}$ a partial
ordering. 
Directly, the map between collections of spatial sets and 
bounds on a given observable described above, restricted
to $\mathcal{C}$, is monotonic. 
If $\mathcal{R}'\geq \mathcal{R}$ then the associated bound for 
$\mathcal{R}'$ is necessarily tighter (bigger or smaller depending 
on the objective) than the bound for $\mathcal{R}$. 
Every refinement $\mathcal{R}'\geq\mathcal{R}$ in $\mathcal{C}$ 
results in a split of multipliers of the optimization Lagrangian, as 
each constraint is decomposed into a set of constraints over the 
matching subregions. 
Therefore, the codomain of the dual function corresponding to 
$\mathcal{R}'$ contains the codomain of the dual function 
corresponding to $\mathcal{R}$, and so the minima (resp. maxima) of 
the dual functions maintains the ordering of $\mathcal{C}$, leading 
to a monotonically tighter bounds, Fig.~1.
Hence, successive division of the spatial regions used in generating
constraints via \eqref{Tformal} indeed yields a well defined 
hierarchy of $\mathbb{T}$ operator bounds, approximating any 
optimization problem with increasingly better accuracy. 
Moreover, less refined solutions are always dual feasible points 
for more refined optimizations, and by evaluating the pointwise 
versions of \eqref{regT} and \eqref{regTR} that would hold under 
complete compliance with the scattering theory for any given dual 
solution $\left|\textbf{T}_{d}\right>$ (resp. $\left\{\left|
\textbf{T}_{d}\right>,~\left|\textbf{R}_{d}\right>\right\}$), as in 
the heat maps of Fig.~1 and Fig.~2, it is possible to assess where 
inconsistency is occurring, and use this knowledge to inform further 
regional decompositions. 

The hierarchy construction amounts, conceptually, to a set of
successively expanded mean-field theories~\cite{soukoulis1977cluster,
plefka1982convergence,jin2016cluster}. 
Traditionally, one of the standard ways in which a mean-field theory 
is constructed is to consider the question of minimizing the Gibb's 
free energy of a system over the collection of all possible 
statistical distributions, parameterized by constraints on its 
moments (expectation values, two-point correlations, etc). 
To make such problems tractable, the form of the distribution is 
simplified in some way (e.g. partitioning the true system into 
effectively interacting spatial clusters) before carrying out a 
local optimization on the resulting (generally
nonconvex) problem. 
The solution, stationary and self-consistent with respect to 
the moments, is called a mean field since to lowest order it 
describes each component of the system as interacting with one other 
``averaged'' body. 
By switching to a scattering description, the need for 
simplification is shifted from the objective to the constraints, but 
the interpretation of the solution is virtually unaltered. 
In either case, it is a field that is self-consistent when 
variations around average values are neglected.

\textit{Bounds}---Generalizing the program given in Ref.~\cite{molesky2020t}, in any order of the hierarchy, the calculation of 
bounds for any net power transfer (scattering) 
objective can be equated with a domain monotonic optimization 
problem described by a sesquilinear Lagrangian.
Grouping the source terms for the constraints and objective,
$\left|\textbf{S}\right>$ and $\left|\textbf{Q}\right>$, together as 
the super source $\left|\underline{\textbf{S}}\right>$
\begin{align}
  \mathcal{L}\left(
  \left\{\alpha_{k}^{\left(1\right)}\right\},
  \left\{\alpha_{k}^{\left(2\right)}\right\}, 
  \left|\textbf{T}\right>
  \right) 
  = 
  \begin{bmatrix}
    \left<\textbf{T}\right| &
    \left<\underline{\textbf{S}}\right|  
  \end{bmatrix}
  \begin{bmatrix}
    \mathbb{Z}^{_{TT}} &
    \mathbb{Z}^{_{T\underline{S}}} \\
    \mathbb{Z}^{_{\underline{S}T}} &
    \textbf{0}
  \end{bmatrix}
  \begin{bmatrix}
    \left|\textbf{T}\right> \\
    \left|\underline{\textbf{S}}\right>
  \end{bmatrix};
  \label{Lgen}
\end{align}
where, supposing an objective of the form 
$\im{\left<\textbf{Q}|\textbf{T}\right>} - 
\left<\textbf{T}\right|\mathbb{O}\left|\textbf{T}\right>$~\cite{
molesky2020t}, the $\mathbb{Z}^{_{-,-}}$ operators are 
the linear couplings (depending on the multipliers) 
between the various fields 
\begin{align}
  -\mathbb{Z}^{_{TT}} &= \mathbb{O} + \sum_{k\in K}\alpha_{k}^{\left(1\right)}\sym{\mathbb{U}\mathbb{I}_{_{\Omega_{k}}}}
  + \alpha_{k}^{\left(2\right)}
  \asym{\mathbb{U}\mathbb{I}_{_{\Omega_{k}}}} 
  \nonumber \\
  &= \mathbb{O} + \sym{\mathbb{U}\mathbb{R}^{\left(1\right)}} + 
  \asym{\mathbb{U}\mathbb{R}^{\left(2\right)}}, 
  \nonumber \\
  \mathbb{Z}^{_{TS}} &= \mathbb{Z}^{_{ST}*} = \sum_{k\in K}
  \frac{\alpha_{k}^{\left(1\right)}}{2}\mathbb{I}_{_{\Omega_{k}}}
  +
  \frac{i\alpha_{k}^{\left(2\right)}}{2}\mathbb{I}_{_{\Omega_{k}}}
  =\frac{\mathbb{R}^{\left(1\right)}}{2} + 
  \frac{i\mathbb{R}^{\left(2\right)}}{2},
  \nonumber \\
  \mathbb{Z}^{_{TQ}} &= \mathbb{Z}^{_{QT}*} = \frac{i}{2}\mathbb{I},
  \label{zMats}
\end{align} 
$\mathbb{R}^{\left(1\right)} = \sum_{k\in K}
\alpha_{k}^{\left(1\right)} \mathbb{I}_{_{\Omega_{k}}}$,
$\mathbb{R}^{\left(2\right)} = \sum_{k\in K}
\alpha_{k}^{\left(2\right)}\mathbb{I}_{_{\Omega_{k}}}$, and
$\left\{\alpha^{\left(1\right)}_{k}\right\}$ and $\left\{
\alpha^{\left(2\right)}_{k}\right\}$ are the sets of Lagrange
multipliers for all symmetric and anti-symmetric constrains
respectively.  For both scattering and radiative Purcell enhancement,
$\mathbb{O} = \asym{\mathbb{V}^{-1\dagger}}$; see
Ref.~\cite{molesky2020t} for further details.  Because optimization is
always formulated in terms of real numbers, the total $\mathbb{Z}$
matrix is Hermitian.

The unique stationary point of $\eqref{Lgen}$ with respect to 
variations in $\left|\textbf{T}\right>$ occurs when $
\left|\textbf{T}\right> = \mathbb{Z}^{_{TT}-1}
\mathbb{Z}^{_{T\underline{S}}}\left|\underline{\textbf{S}}\right> 
=\mathbb{Z}^{_{TT}-1}\left(\mathbb{Z}^{_{TS}}\left|
\textbf{S}\right> + \mathbb{Z}^{_{TQ}}
\left|\textbf{Q}\right>\right),$
and so the dual of \eqref{Lgen}, $\mathcal{G} = 
\text{max}_{\mathcal{F}}~\mathcal{L}$ where the domain 
$\mathcal{F}$ is set by the criterion that $\text{max}~\mathcal{L}$ is finite, is
\begin{equation}
  \mathcal{G} =
  -\left<\underline{\textbf{S}}\right|
  \mathbb{Z}^{_{\underline{S}T}}
  \mathbb{Z}^{_{TT-1}}
  \mathbb{Z}^{_{T\underline{S}}}
  \left|\underline{\textbf{S}}\right>.
  \label{gSol}
\end{equation}
Note that a set $\left\{\left<\alpha^{\left(1\right)}_{k},
\alpha^{\left(2\right)}_{k}\right>\right\}$ lies within $\mathcal{F}$
so long as $\mathbb{Z}^{_{TT}-1}$ is positive
definite~\cite{strang2007computational}.

Because the dual problem is convex and gradients of the dual 
reproduce the constraints set by \eqref{regT}, if a stationary point 
within the feasibility region is found, then strong duality holds 
and the maximum of the objective is $\im{\left<\textbf{Q}|\textbf{T}
\right>} - \left<\textbf{T}\right|\mathbb{O}
\left|\textbf{T}\right>$, with $\left|\textbf{T}\right>$
determined as above using the multiplier values set by the
simultaneous zero point of the gradients (constraints). 
If no such point exists in $\mathcal{F}$, then the unique minimum 
value of $\mathcal{G}$ attained on the boundary, caused by 
$\mathbb{Z}^{_{TT}}$ becoming semi-definite, remains a bound on the 
optimization problem.

\textit{Example}---As an initial exploration of $\mathbb{T}$ operator
constraint hierarchies, under \eqref{regT}, Fig.~2 illustrates how the
introduction of spherically symmetric (concentric) shell clusters
alters bounds on Purcell enhancement (radiative emission from a
dipolar current source in the vicinity of a structured medium
normalized to emission in vacuum) compared to past global-conservation
arguments~\cite{molesky2020t,gustafsson2019upper}.  Even with a choice
of relatively simple clusters, which leads to considerable numerical
simplifications, but also limit the extent to which violation can be
localized, two promising trends are observed.  First, the number of
material and domain size combinations displaying resonant response
characteristics, as qualified by a roughly inverse scaling between
radiative enhancement and material loss $\text{Im}\left[\chi\right]$,
is reduced.  To the best of our knowledge, no compact single
dielectric material device design exhibits such behavior for values of
$\text{Im}\left[\chi\right]$ comparable to those considered here,
leading to potentially gigantic gaps between calculated limits and the
findings of density optimization, Fig.~2 (a).  So long as strong
duality is present, which holds for all of (a) and (b) below
$R/\lambda \approx 0.3$, the dependence of the cluster bounds on
$\text{Im}\left[\chi\right]$ is greatly suppressed; and when a
relation between these quantities is observed for
$\text{Im}\left[\chi\right]\lesssim 10^{-2}$, the corresponding
optimal polarization field, $\left|\textbf{T}\right>$, implies large
local constraint violations, as shown by the spatial color maps of the
real part of \eqref{regT} included below Fig.~2 (c).  Similarly,
Fig.~2 (b) and (c), for separations as small as $d/\lambda = 10^{-2}$
and domains as large as $R/\lambda = 1$, cluster-bound values are
substantially smaller (often an order of magnitude or more) for strong
metallic and dielectric materials, $|\text{Re}\left[\chi\right]| \gg
5$.  Nevertheless, the limits remain several orders of magnitude
larger than what is observed in devices discovered by density
optimization when such values of $\chi$ are supposed, suggesting
further room for improvement.

This work was supported by the National Science Foundation under
Grants No. DMR 1454836, DMR 1420541, DGE 1148900, 
EFMA 1640986, the Cornell Center for Materials Research MRSEC 
(award no. DMR1719875), and the Defense
Advanced Research Projects Agency (DARPA) under Agreement
No. HR00112090011. 
The views, opinions and/or findings expressed
herein are those of the authors and should not be interpreted as
representing the official views or policies of any institution. 
The authors thank Cristian L. Cortes for useful discussion. 
\bibliography{esaLib}
\end{document}